\documentclass[aps,amsfonts,preprint,showpacs,floatfix]{revtex4}

\usepackage{graphicx}
\usepackage{amsmath}
\usepackage{bm}

\begin{document}

\title{Kinetic equations for the hopping transport and spin relaxation in random magnetic field.}
\author{A. V. Shumilin$^1$, V.V. Kabanov$^2$}

\affiliation{$^1$A.F.Ioffe Physico-Technical Institute, St.-Petersburg 19
4021, Russia.\\
$^2$Department for Complex Matter, Jozef Stefan Institute, 1001 Ljubljana, Slovenia.}

\begin{abstract}
We derive the kinetic
equations for the hopping transport that take into account electron spin and the possibility of double occupation.
In the Ohmic regime the equations are reduced to the generalized Miller-Abrahams resistor network. We apply these equations to the problem of the magnetic moment
relaxation due to the interaction with the random hyperfine fields. It is shown that in a wide range of parameters the relaxation rate is governed by the hops with
the similar rates as spin precession frequency. It is demonstrated that at the large time scale
spin relaxation is non-exponential. We argue that the non-exponential relaxation of the magnetic moment is related to the spin of electrons in the slow-relaxing traps. Interestingly the traps can significantly influence the spin relaxation in the infinite conducting cluster at large times.
\end{abstract}


\maketitle

\section{Introduction}

In recent years investigations of the spin phenomena in hopping
transport have gone through a vigorous revival. In particular,
the increasing interest to this problem is related to the discovery of a
strong spin-valve effect in organic semiconductor devices\cite{dediu2002,vardeny2004}.
There is a consensus that the conductivity in these devices is determined
by hopping polarons\cite{Basler_polarons}. Such devices show a number of spin-related phenomena,
including the spin-valve effect itself that are not understood. These experimental results were followed by a number of
theoretical investigations of the hopping conduction \cite{Pasveer,Cottaar} including fluctuations of conductivity \cite{Masse,Lukyanov} and most important the spin phenomena like
magnetoresistance \cite{HF1,HF2,HF3,ADK,Bobbert} and the relaxation of magnetic
moment\cite{ShklovskiiR,Dmitriev,Kavokin,HF4,HF5, Yu-SO, Yu-SO2}.

Most of these theoretical studies are based on semi-qualitative concepts with the lack of solid
theoretical proof. Although in some cases (for example in the case of strong Coulomb interaction)
the self consistent theory of hopping transport does not exist, there is a way to make theory of
hopping transport self consistent. This theory is described in Ref.\cite{Bryksin-Book}. Up to now
the theory includes the electron spin only in the limit of the small electron
density\cite{Bryksin-JETP,Bryksin-PRB,Bryksin-PRB0}.

Here we develop the generalization of this theory to include spin and arbitrary probability of
the site occupation (with possible double occupation). We argue that the self consistent theory of
hopping transport is necessary in order to test the semi-qualitative concepts \cite{HF1,HF2,HF3}  and to understand the electron and the spin transport in organic semiconductors.

Our starting point is the general Hamiltonian that describe a system of localized sites with the possibility of hopping due to electron-phonon interaction. Then we derive general kinetic equations that describe both the charge and the spin transport in  hopping media. We argue that these equations are a useful tool to study hopping transport phenomena.

In the present paper we apply these kinetic equations to the problem of the spin relaxation due to random on-site magnetic fields in a system with the positional disorder.  The understanding of the spin relaxation is clearly important for the theory of the spin-related transport phenomena like spin-valve magnetoresistance. Up to now the most studied mechanism of the spin relaxation in hopping transport is the spin-orbit interaction. The theory of this relaxation mechanism is discussed in Refs.\cite{ShklovskiiR,Dmitriev,Kavokin, Yu-SO, Yu-SO2}. In Ref.\cite{ShklovskiiR} the basic
understanding of spin relaxation for the hopping transport due to spin-orbit interaction was
formulated. In Ref.\cite{Dmitriev} the results of \cite{ShklovskiiR} were significantly expanded.
It was pointed out \cite{Dmitriev} that exponentially broad distribution of hopping rates plays
determinant role in the relaxation. However this distribution was considered semi-quantitatively. In Ref.\cite{Kavokin} it was proposed that in the case
of hopping conduction the spin relaxation due to the spin-orbit interaction may be significantly enhanced by the exchange interaction.

While the spin-orbit interaction is supposed to dominate the spin relaxation in non-organic semiconductors with hopping conductivity the situation in the organic materials may be different. In organic semiconductors the spin-orbit interaction is substantially suppressed \cite{Yu-SO, Yu-SO2, SO_inO}
and another mechanisms like hyperfine interaction can govern the spin relaxation. The theoretical description of this mechanisms is different because the spin-orbit interaction does not affect the spin of a localized electron and manifest itself only in the spin rotation during the hop. On the other hand the hyperfine interaction leads to the appearance of random effective on-site magnetic fields that rotate electron spins even without hops. Note that in the organic spin-valve devices there is the another source of random fields unique to these case. The finite roughness of the contacts leads to a leakage of the magnetic field from the ferromagnetic contacts to the organic layer \cite{Fringe}. This fringe magnetic field can be a source of additional mechanism of the spin relaxation. In terms of theoretical description it is added to the hyperfine field and also rotate spins on the localized sites.

Recently Harmon and Flatte \cite{HF4, HF5} proposed an interesting approach to the spin relaxation based on the waiting time distribution. They considered both the spin-orbit and the hyperfine relaxation mechanisms. However their approach does not take into account the Pauli principe and is applicable only to the limit of the small electron concentrations.  Also Refs.\cite{HF4, HF5} consider only the energy disorder and the positional disorder was neglected.

The derived kinetic equations allow us to describe spin relaxation with any site occupation probability. In the present work we apply them to the problem with the positional disorder and neglect the energy disorder. We show that even in this case there are several phenomena in the spin relaxation that were not discussed previously. Most important of them is perhaps the non-homogeneous character of the spin relaxation. In some cases the relaxation of spin of the electrons important for conduction is substantially different from the relaxation of the average spin of the system.

The paper is organized as follows. In section \ref{sect_kin1} we discuss the usual way of describing hopping conduction and its justification in terms of kinetic equations. In section \ref{sect_kin2} we generalize approach \cite{Bryksin-Book} to include electron spin and derive general form of kinetic equation. In section \ref{sect_kinlin} we linearize the kinetic equations and derive generalized version of Miller-Abrahams resistor network. Finally in section \ref{sect_relax} we use the kinetic equations to describe the spin relaxation in the hopping conduction regime.

\section{Kinetic equations for hopping transport}
\label{sect_kin1}

The usual approach to the theoretical description of the hopping transport starts with the
introduction of the hopping rates between pairs of sites \cite{Efr-Sh}. The hopping rate from site $i$ to site $j$ is defined as
\begin{equation} \label{Gamma}
\Gamma_{ij} \propto \left| I_{ij} \right|^2 f^{(i)} (1-f^{(j)}) P_{ph}(\Delta E_{ij}).
\end{equation}
Here $I_{ij} \propto \exp(-r_{ij}/a)$ is the overlap integral between sites $i$ and $j$, $r_{ij}$
is the distance between these sites, $a$ is the localization radius. $f^{(i)}$ and $f^{(j)}$ are the
occupation probabilities of sites $i$ and $j$. $\Delta E_{ij}$ is the difference of energies of
states $i$ and $j$. $P_{ph}(\Delta E_{ij})$ is the part of hopping rate related to the number of
phonons, involved in the hopping. It is equal to $N_{ph} + 1$ when $\Delta E_{ij}>0$, i.e., when
the hopping occurs with the phonon emission and to $N_{ph}$ otherwise. Here $N_{ph}$ is the number of phonons with energy $|\Delta E_{ij}|$.

Then the current between sites $i$ and $j$ is introduced as
\begin{equation} \label{current}
J_{ij} = -e(\Gamma_{ij} - \Gamma_{ji}).
\end{equation}
$J_{ij} =0$ when the external electric field is absent. In a weak electric field the current
follows the Ohm law $J_{ij} = U_{ij}/R_{ij}$ with effective resistor voltage $U_{ij} = \Delta\varphi_{ij} - \Delta\mu_{ij}/e$ corresponding to the shift of electrochemical
potential between sites $i$ and $j$. The resistance $R_{ij}$ is defined as
\begin{equation}\label{resist}
R_{ij} = \frac{k_B T}{e^2 \Gamma_{ij}^{(0)}},
\end{equation}
where $\Gamma_{ij}^{(0)}$ is the hopping rate between sites $i$ and $j$ without the external
field, $k_B$ is the Boltzmann constant.

As a result the real system with the hopping conductivity is replaced by a network
of classical
resistors. This network can be treated with percolative methods. In this case one
finds the threshold resistance that allows the percolation to the macroscopic distances.
In the case when the distribution of resistances $R_{ij}$ is exponentially broad, this
threshold resistance governs the conductivity of the whole system.

 Without Coulomb interaction this approach can be consistently derived in terms of kinetic equations \cite{Bryksin-Book}. It is the starting point of many theoretical studies and works quite well in many cases. However it has some problem when the hopping conductance becomes dependent on the electron spin because  spin is not included in the equations (\ref{Gamma} - \ref{resist}). Let us discuss how this scheme can be modified in order to include electron spin and how it was altered in previous studies.

The essential part of the discussed treatment is the percolation theory that is a conventional way to study analytically dc current in a hopping system with a broad distribution of hopping rates.  As long as the hopping rates depend on spin and electron spins are not totally polarized the theory should also contain some sort of averaging over possible spin direction.
There are at least two possible orders of this procedures: one can first do the spin averaging and then calculate the percolation parameters or calculate the percolation threshold first and then average over spin (or make some more sophisticated approach to relating these two procedures). Different existing studies apply different order of percolation and spin averaging.

In Ref. \cite{Matveev} the percolation was calculated over resistors network that considered both probabilities of having spin up and spin down electron on each site. These probabilities depend on the magnetic field when Zeeman energy is of the order of $kT$ leading to the magnetoresistance. The percolation with rates averaged over spin directions was also implicitly considered in \cite{SS} and \cite{ZhaoSpivak} where it was noted that electron spins should be freezed in order to have negative interference magnetoresistance (otherwise the averaging over spin directions kills the effect).

The another relation between percolation and spin averaging was considered in Refs. \cite{HF1,HF2,HF3}. In these articles it was allowed double-occupation of the sites but only for electrons with antiparallel spins (more exactly: in singlet spin state). This assumption agrees with \cite{Matveev} but the following approach is different.

  The  approach in \cite{HF1,HF2,HF3} starts  with momentary site occupations and momentary spin projections. Then the percolation is considered in terms of effective concentration of sites allowed for the hop of a given electron. Double occupied sites are always excluded from the percolation (disregarding the fact that they may be empty in following moments of time). Single occupied sites are included in the percolation when the spin on the site is antiparallel to the spin of the hopping electron. If electron spins are parallel the site is included in percolation with some probability $p(H)$ reflecting that the spin flip can occur faster than the hopping of an electron to a distant site.

With $p=0$ the percolation is calculated before spin averaging. Finite $p(H)$ leads to a more sophisticated relation between percolation and spin averaging. The dependence of $p$ on magnetic fields leads to the magnetoresistance.

The approach different from previous two was proposed by Osaka \cite{Osaka}. The percolation in the model of the resistor network was considered. The spin-flip process was considered as an additional resistor connected in parallel to a normal one.

We want to underline that the discussed approaches are not equivalent and lead to different physical results. For example in the approach of \cite{Matveev} the concentration of sites that participate in percolation is the full concentration of sites as long as there is some probability for any site to be allowed for the hop. In \cite{HF1,HF2,HF3} at least double occupied sites are always excluded from the percolation.

To understand what is the correct procedure of spin introduction into the percolation theory let us consider the procedure of the consistent derivation of the approach (\ref{Gamma}). This procedure is described by Bottger and Bryksin \cite{Bryksin-Book}, however we remake it here in a slightly different way to make its generalization easier. Note that the discussion in Ref. \cite{Bryksin-Book} is focused on the case of low occupation numbers and  it is stated that that the result for any occupation number is similar if the Hartree decoupling is used.  We make our theory with Hartree expansion from the beginning..

We start with the hopping Hamiltonian after polaron transformation, which is well known in the polaron transport theory\cite{Bryksin-Book}.
\begin{equation}\label{H_pol}
H = H_0 + H_{hop}, \quad H_0 = \sum_i (\varepsilon_i-E_p) a_i^+ a_i + \sum_q \hbar \omega_q \left( b_q^+ b_q + \frac{1}{2} \right),
\end{equation}
$$
H_{hop} = \sum_{ij} t_{ij} a_i^+ a_j \hat{\Phi}_{i,j}.
$$
$$
\hat{\Phi}_{i,j}=\exp{\Bigl \{ \sum_{\bf q} [b_{\bf q}^+M_{\bf q}^*(e^{-i{\bf q}{\bf r}_i}-e^{-i{\bf q}{\bf r}_j})/\hbar\omega_q-h.c.] \Bigr \}}
$$
Here $a_i$ is the electron  annihilation operator on site $i$, $b_{\bf q}$ is the annihilation
operator for a phonon with the wave-vector ${\bf q}$ and with frequency $\omega_q$. We take
into account an energy disorder therefore each site has a random energy $\varepsilon_i$.
$t_{ij}$ are the overlap integrals between the sites. $t_{ij}$ are much smaller than
the differences in random energies $\varepsilon_i - \varepsilon_j$. $M_{\bf q}$ is the
electron-phonon interaction matrix element $M_{\bf q} = M_{-{\bf q}}^*$. ${\bf r}_i$ is the position of the
site $i$, $E_p=\sum_{\bf q} |M_{\bf q}|^2/\hbar \omega_q$ is the polaron binding energy.

The conventional derivation of the kinetic equation \cite{Silin} starts with the full description
of the system (with Liouville equations if the system is classical or with Hamiltonian if
the system is quantum). Then the system is divided into the parts that interact weakly or
rarely. We divide the system into the set of noninteracting sites,
described by the Hamiltonian $H_0$, and weak interaction of different sites, described by
the Hamiltonian $H_{hop}$. In order to describe the system with the set of kinetic equation
we have to assume that the eigenstates of the Hamiltonian
$H_0$ are well defined and therefore $H_{hop}$ is much smaller  than $H_0$.

In zero order over $H_{hop}$ the full density matrix of the system is diagonal in terms of electron filling numbers $n_i$.
The theory of hopping conduction near Fermi level (that we are interested in) is based on the Hartree-Fock decoupling for the density matrix \cite{Bryksin-Book}. With this decoupling the full density matrix can be expanded as a product
of the single site density matrices.
\begin{equation}\label{matr-all}
\rho_{n_1, n_2,...,n_{N}}^{n_1',n_2',...n_N'} = \rho_{n_1}^{n_1'(1)}\cdot \rho_{n_2}^{n_2'(2)} \cdot ... \cdot \rho_{n_N}^{n_N'(N)}.
\end{equation}
Here we used the basis of the filling numbers. The set of the system states that has definite
filling numbers $n_{i}$ is the complete set of states of the system. As usual the density
matrix has two indexes each of them corresponding to one of the states of the complete set. The upper indexes in round brackets
correspond to a number of site.

The single site density matrix $\rho_{n_i}^{n_i'(i)}$ has only two indexes $n_i=0,1$ that are
the possible filling numbers of site $i$. Without $H_{hop}$ this matrix is defined as:
\begin{equation} \label{matr1}
\rho^{(i)} = \left(
\begin{array}{cc}
\rho_1^{1(i)} & \rho_1^{0(i)} \\
\rho_0^{1(i)} & \rho_0^{0(i)}
\end{array}
\right)
=
\left(
\begin{array}{cc}
f^{(i)} & 0 \\
0 & 1-f^{(i)}
\end{array}
\right)
\end{equation}
where $f^{(i)}$ is the probability for site $i$ to have an electron.

Here we will use the interaction representation, therefore $\rho_{n_i}^{n_i'(i)}$ does not
depend on time without perturbation. In the first perturbation order we obtain
\begin{equation} \label{rho1}
\frac{d \rho^{(i)}}{dt} = \frac{1}{i\hbar}\sum_j {\rm Tr}_j \left[ (\widetilde{H}_{hop})_{ij}, \rho^{(ij)} \right], \quad \rho^{(ij)} = \rho^{(i)}\rho^{(j)} + d^{(ij)}.
\end{equation}
Here $(\widetilde{H}_{hop})_{ij}$ is the part of $H_{hop}$ corresponding to the hops between
sites $i$ and $j$ in the interaction representation. $\rho^{(ij)}$ is the two-site density matrix.
We divide $\rho^{(ij)}$ into the product of the one-site density matrices and a small off-diagonal (in terms of electron filling numbers) correlated part
$d^{(ij)} \propto H_{hop}$.
The contribution of the product $\rho^{(i)}\rho^{(j)}$  corresponds
to the mean-field correction to the energy $\varepsilon_i$ and is usually neglected.
The time evolution of $\rho^{(i)}$ is governed by the off-diagonal part $d^{(ij)}$.

To obtain the
equation for $d^{(ij)}$ one should write the equation for the two site density matrix
\begin{equation}\label{rho2}
\frac{d \rho^{(ij)}}{dt} = \frac{1}{i\hbar}\left[(\widetilde{H}_{hop})_{ij}, \rho^{(ij)} \right] + \frac{1}{i\hbar}\sum_k {\rm Tr_k} \left[(\widetilde{H}_{hop})_{ik} + (\widetilde{H}_{hop})_{kj}, \rho^{(ijk)} \right].
\end{equation}
Note that  $d^{(ij)}$ enters equation (\ref{rho1}) with the coefficient $\propto H_{hop}$.
The right hand side of Eq.(\ref{rho2}) also contains $H_{hop}$. Here we neglect all powers of $H_{hop}$ higher than 2. Therefore we replace all many-site density matrices by the products of the one-site density matrices in the right hand side of Eq.(\ref{rho2}). As a result we obtain the expression for $d^{(ij)}(t)$:
\begin{equation}\label{d2}
d^{(ij)}(t) = \frac{1}{i\hbar} \int_{-\infty}^t \left[(\widetilde{H}_{hop})_{ij}(t'), \rho^{(i)}(t')\rho^{(j)}(t') \right] dt'.
\end{equation}
To get the kinetic equation we assume that the correlated part $d^{(ij)}(t)$ decays much faster in comparison with the time scale on which the one-particle density matrix changes. Therefore the one-site density matrices are out of the integral in Eq.(\ref{d2}). This assumption allows to get the final equation for the one-site density matrix:
\begin{equation}\label{rho-gen}
\frac{d \rho^{(i)}}{dt} = -\frac{1}{\hbar^2}\sum_j \left\langle {\rm Tr}_j \left[(\widetilde{H}_{hop})_{ij}(t),
\left[ \int_{-\infty}^t (\widetilde{H}_{hop})_{ij}(t') dt', \rho^{(i)}(t) \rho^{(j)}(t)  \right]
 \right] \right\rangle_{ph}.
\end{equation}
$H_{hop}$ contains not only electron but also phonon operators. Therefore we should average the right-hand side of this equation over phonons. Substituting Eqs.(\ref{H_pol}) and (\ref{matr1}) into (\ref{rho-gen}) we obtain the following result:
\begin{equation} \label{df/dt}
\frac{d f^{(i)}}{dt} = \sum_j W_{ji} f^{(j)} (1-f^{(i)}) - W_{ij} f^{(i)} (1-f^{(j)}),
\end{equation}
where the hopping rate in the limit $M_q/\hbar\omega_q\ll1$ has the form:
\begin{equation} \label{W}
W_{ji} = \frac{2\pi}{\hbar} \sum_{\bf q} \frac{t_{ij}^2}{\hbar^2\omega_q^2}\left|M_{\bf q} \right|^2\left|e^{i\bf{qr}_i} - e^{i\bf{qr}_j} \right|^2 \left[ (N_{ph}+1) \delta(\varepsilon_i - \varepsilon_j + \hbar\omega_q) + N_{ph}\delta(\varepsilon_i - \varepsilon_j - \hbar\omega_q) \right].
\end{equation}
The hoping rates in the limit of $M_q/\hbar\omega_q\ge1$ are derived in\cite{Bryksin-Book}

It is the kinetic equation for the spinless electrons. In order to derive the dc current
one should consider the steady state solutions of these equations $df_i/dt=0$ and then introduce a small electric field. As a result the
equation yields the standard formula for the current (\ref{current}) where $\Gamma_{ij} = W_{ij} f^{(i)} (1-f^{(j)})$.

Let us discuss the applicability conditions of the kinetic equation. First of all it relies on the  smallness of the intersite correlations $d^{(ij)}$. In the discussed problem it is closely related to the Hartree decoupling. It is valid when the system is close to equilibrium \cite{Landau-Kin}. However far from equilibrium the site occupation may be correlated, therefore $\langle a_i^+a_i a_j^+a_j \rangle \ne \langle a_i^+a_i \rangle \langle a_j^+a_j \rangle$. In this case the system cannot be described with a closed equation for $f_i$. Real systems, however can have other sources of these correlations (even in the equilibrium). The most known of them is the Coulomb interaction. The problem of the Coulomb interaction in hopping conductivity is rather long-standing. If the Coulomb interaction between neighboring sites is comparable with the random energies, the correlations can appear even without $H_{hop}$. In that case the applicability of the  kinetic equation becomes questionable even in the spinless case. The standard answer for this problem is that the kinetic equation is still applicable but the density of single electron states should be considered taking into account the Coulomb interaction. It leads to the formation of the Coulomb gap near Fermi level. Equation (\ref{df/dt}) with the Coulomb gap was successfully used to obtain the well-known Efros-Shklovskii temperature dependance of conductivity \cite{Efr-Sh}, however the other consequences of Coulomb interaction are still under discussion.

Another important simplification is neglecting of all the high-order terms of $H_{hop}/\varepsilon_i$. It is known that these higher order terms may lead to the phenomena that are important for the hopping conductivity.  Recently in Refs. \cite{Kavokin, Yu} it was argued that intersite exchange effects may be important for the spin relaxation and the spin transport. Another example is the sub-barrier scattering that is responsible for the linear negative magnetoresistance in semiconductors with the variable-range hopping conductivity \cite{SS}.  Although these phenomena are important, we believe that it is more important to understand low-order physics before consider these high-order phenomena.

\section{Kinetic equation with spin and possibility of double occupation}
\label{sect_kin2}

Up to this moment our results repeat the traditional scheme \cite{Efr-Sh}, at least when dc
current is under consideration. The advantage of our approach is that it may be easily
generalized for the case when electrons have spins and can double occupy a site.

Equation (\ref{rho-gen}) does not depend on the exact structure of the one site density matrix
corresponding to spinless electrons. It relies only on the kinetic equation assumptions that
are valid in general case. Therefore in order to generalize our theory we have to include spin
into the Hamiltonian, define the structure of one-site density matrix in general case, and
perform the calculations that are analogous to the derivation of Eq.(\ref{df/dt}).

Here we consider the case when the electron spin conserves during the hopping (the under-barrier spin rotation is discussed in section \ref{sect_SO}). However we include small
on-site spin Hamiltonian $H_{S}$ that describes  rotation of the spin over the effective local
magnetic field  $H_{S} = \mu_b g \sum_{i} {\bf H}_i \widehat{\bf s}_i$, where ${\bf H}_i$ is the local effective magnetic field and $\widehat{{\bf s}}_i$ is the operator of spin on site $i$.
\begin{equation}
H = H_0 + H_S + H_{hop}.
\end{equation}
$$
H_0 = \sum_{i,s} \varepsilon_i a_{is}^+ a_{is} + \sum_i U a^+_{i+}a_{i+}a^+_{i-}a_{i-}; \quad H_S = \sum_i H_{S,i}
$$
Here $s$ is the spin index that can have two values ``$+$" and ``$-$". $U$ is the Hubbard energy. The introduction of the term $\sum_i U a^+_{i+}a_{i+}a^+_{i-}a_{i-}$ corresponds to the following model. We allow double occupation of the site  but consider that other excited orbital
states at the same site have very large energies. So far both electrons on a double-occupied site have the same coordinate wave-function and their spins should form a singlet state. Double occupation of a site with two electrons in the triplet spin state is not allowed.
 $H_{S,i} = \mu_b g {\bf H}_i \widehat{\bf s}_i$ acts only on electrons on site $i$ and conserve the filling number of this site, i.e., it can only rotate the spin on a single-occupied site.
\begin{equation}\label{H_hop_sp}
H_{hop} = \sum_{ijs} t_{ij} a_{is}^+ a_{js} \hat{\Phi}_{i,j},
\end{equation}
where $\hat{\Phi}_{i,j}$ is defined after Eq.(\ref{H_pol}). The hopping part of the Hamiltonian (\ref{H_hop_sp}) conserves the spin.

The one-site density matrix in the representation of filling numbers $\rho_{i_-,i_+}^{i_-',i_+'}$ now contains four indexes and have 16 matrix elements. We however will use another representation in this section. The complete set of states for one site with possibility of double occupation has four states that can be selected as
\begin{equation} \label{4states}
\left|0 \right>, \, \left|+ \right>, \, \left|- \right>, \, \left|2 \right>.
\end{equation}
Here $\left|0 \right>$ is the empty site, $\left|2 \right>$ is the double occupied site, $\left|+ \right>$ and $\left|- \right>$ are the states of the site when it has one electron with spin up and down respectively. So one can write one-site density matrix with only two indexes, each of them can have any of four values described in (\ref{4states}).

In zeroth order over $H_{hop}$ we have only matrix elements of the one-site density matrix that conserve the filling number. For the spinless electrons there were only two such elements. Now we have six of them.
\begin{equation}\label{5el}
\rho_0^0, \, \rho_{+}^+, \, \rho_+^-, \, \rho_-^+, \, \rho_-^-, \, \rho_2^2.
\end{equation}
Only these 6 matrix elements appear in the kinetic equation.

Along with the density matrix elements (\ref{5el}) it is sometimes useful to consider another set of 6 numbers, that are linear combinations of the
matrix elements (\ref{5el}), in order to describe the state of the site. These numbers allow us to
track directly  the occupation number and the mean value of the magnetic moment of the site.
The transition to these numbers was proposed by Bryksin (without double occupation
probability)\cite{Bryksin}.
\begin{equation}\label{5bryk}
\begin{array}{ll}
f_0 = \rho_0^0, & f_1 = \rho_+^+ + \rho_-^-,\\
M_z = \rho_+^+ - \rho_-^-, & M_x  = \rho_-^+ + \rho_+^-, \\
M_y = -i\rho_+^- + i \rho_-^+, \qquad & f_2 = \rho_2^2.
\end{array}
\end{equation}
Here $f_0$, $f_1$ and $f_2$ are the probabilities for a site to have 0, 1 and 2 electrons
respectively. $M_\alpha$ is the mean magnetic moment of the site in the direction $\alpha$.

The generalized equation (\ref{rho-gen}) in these notations has the following form:
\begin{equation}\label{kin-gen}
\frac{d \rho_x^{(i)}}{d t} - {\cal S}_{xy}(i) \rho_{y}^{(i)} = \sum_j W_{xyz}(ij)\rho_y^{(i)}\rho_z^{(j)}.
\end{equation}
Here $x$, $y$ and $z$ have 6 possible values, $\rho_x$ correspond to some filling number
probabilities or mean projections of magnetic moment defined in (\ref{5bryk}). The term
${\cal S}_{xy}(i) \rho_{y}^{(i)}$ corresponds to the action of the spin part of the Hamiltonian
$H_S$.
$$
{\cal S}_{M_\alpha,M_{\beta}} =  \frac{\mu_b g}{\hbar} \epsilon_{\alpha\beta\gamma}  H_{\gamma}^{(i)} = \epsilon_{\alpha\beta\gamma} {\cal H}_{\gamma}^{(i)}.
$$
Here $\epsilon_{\alpha\beta\gamma}$ is  Levi-Civita symbol, $\mu_b$ is the Bhor magneton and $g$ is the g-factor.
It correspond to the precession of
the local magnetic moment $d {\bf M} (i)/dt = (\mu_b g /\hbar)\left[ {{\bf M}(i), \bf H}(i) \right]$.
We also introduced here a renormalized local magnetic field measured in units of frequency $\vec{\cal H}^{(i)} = (\mu_b g/\hbar){\bf H}^{(i)}$

The important assumption related to Eq. (\ref{kin-gen}) is that the on-site
Hamiltonian can be treated independently from $H_{hop}$. It is valid when
$H_S \delta t_{Mark}/ \hbar \ll 1$,
where $\delta t_{Mark}$ is the characteristic decay time of the correlation $d_{ij}$. (In
the opposite limit the kinetic equation becomes non-Markovian). We also did not include
the Zeeman energy when the
averaging over phonon states Eq. (\ref{rho-gen}) is discussed. This assumptions are
justified when $H_S$ is small compared to $\varepsilon_i - \varepsilon_j$
and $kT$

For larger magnetic fields especially when Zeeman energy becomes larger than temperature the kinetic equations in the present form
are valid only when the site magnetization and the magnetic field are oriented along one axis (in that case $H_S$ does not lead to the magnetization precession). In that case the site energies in the equation should include the Zeeman energy. If the on-site magnetic field is
large and is oriented along different axes the phonon averaging can lead to more complex equations. This case is however out of the scope of the present work.

The term $W_{xyz}(ij)\rho_y^{(i)}\rho_z^{(j)}$ is the ``collision integral"
\begin{equation}\label{colint}
-\frac{1}{\hbar^2} \left\langle {\rm Tr}_j \left[(\widetilde{H}_{hop})_{ij}(t),
\left[ \int_{-\infty}^t (\widetilde{H}_{hop})_{ij}(t') dt', \rho^{(i)}(t) \rho^{(j)}(t)  \right]
 \right] \right\rangle_{ph}
\end{equation}
represented in the notations (\ref{5bryk}). Each of the indexes $x$, $y$ and $z$ can have 6 different values, so there are $6^3$ matrix elements $W_{xyz}$ and their calculation is rather cumbersome. Using the trick described in the Appendix we derive the following set of kinetic equations:
\begin{equation}\label{kin_f0}
\frac{df_0^{(i)}}{dt} = \sum_{j\ne i} W_{ij}f_1^{(i)}f_0^{(j)} + \frac{W_{ij}^{-U}}{2}\left[f_1^{(i)} f_1^{(j)} - M_{\alpha}^{(i)} M_{\alpha}^{(j)}\right] -
\end{equation}
$$
-W_{ji}f_0^{(i)} f_1^{(j)} - 2W_{ji}^{+U} f_0^{(i)} f_2^{(j)}.
$$
\begin{equation}\label{kin_f}
\frac{df_1^{(i)}}{dt} = \sum_{j \ne i} W_{j i} f_1^{(j)} f_0^{(i)} + 2 W_{ji}^{+U} f_2^{(j)} f_0^{(i)} + W_{ij} f_2^{(i)}f_1^{(j)} + 2W_{ij}^{+U}f_2^{(i)}f_0^{(j)}-
\end{equation}
$$
- \frac{W_{ij}^{-U}+W_{ji}^{-U}}{2}\left[f_1^{(i)} f_1^{(j)} - M_\alpha^{(i)} M_{\alpha}^{(j)}\right] - W_{ij}f_1^{(i)}f_0^{(j)} - W_{ji}f_1^{(i)}f_2^{(j)};
$$
\begin{equation} \label{kin_M}
\frac{d M_{\alpha}^{(i)}}{dt} + \epsilon_{\alpha\beta\gamma} M_{\beta}^{(i)}{\cal H}_{\gamma}^{(i)} = \sum_{j\ne i}W_{ji} M_\alpha^{(j)}f_0^{(i)} + W_{ij}M_{\alpha}^{(j)}f_2^{(i)} +
\end{equation}
$$
+ \frac{W_{ij}^{-U}+W_{ji}^{-U}}{2} \left[M_{\alpha}^{(j)} f_1^{(i)} - f_1^{(j)}M_{\alpha}^{(i)}\right]  - W_{ij}f_0^{(j)}M_\alpha^{(i)} - W_{ji}f_2^{(j)}M_\alpha^{(i)}.
$$
\begin{equation}\label{kin_f2}
\frac{d f_2^{(i)} }{d t} = \sum_{j\ne i} W_{ji}f_1^{(i)}f_2^{(j)} + \frac{W_{ji}^{-U}}{2}\left(f_1^{(i)}f_1^{(j)} - \sum_{\alpha} M_{\alpha}^{(i)} M_{\alpha}^{(j)}\right) -
\end{equation}
$$
- W_{ij}f_2^{(i)}f_1^{(j)} - 2 W_{ij}^{+U} f_2^{(i)}  f_0^{(j)}.
$$

Here we have introduced the hopping rates $W_{ij}^{+U}$ and $W_{ij}^{-U}$. One can note that
$W_{ij}$ defined in (\ref{W}) depends on the energy difference $\varepsilon_i - \varepsilon_j$.
However if one of the initial and the final state of the hop corresponds to the upper Hubbard band,
the actual energy $\varepsilon_i + U_h$ or $\varepsilon_j + U_h$ should be used. Therefore
$W_{ij}^{+U}$ is the hopping rate $W_{ij}$ where $\varepsilon_i$ is substituted by
$\varepsilon_i + U_h$ and $W_{ij}^{-U}$ is $W_{ij}$ with $\varepsilon_j$ substituted with
$\varepsilon_j + U_h$.

\subsection{Spin-orbit couplings}
\label{sect_SO}

The kinetic equations Eqs.(\ref{kin_f0}-\ref{kin_f2}) were derived with the approximation that
the electron spin is conserved during the hop. This approximation is not sufficient when the spin-orbit
interaction is essential for the kinetics. Although we do not want to discuss the role of the
spin-orbit interaction in details we briefly outline the procedure of the
inclusion of the spin-orbit interaction in the kinetic equations in this section.

The possibility of inclusion of the spin-orbit interaction into kinetic equation for hopping
conductivity was discussed in \cite{Bryksin-PRB0, Bryksin-JETP,Bryksin-PRB}, where corresponding
kinetic equations were derived in the limit of low occupation numbers. The spin-orbit interaction
leads to a precession of the electron spin during the under-barrier motion. It is important that
for a given pair of sites $i$ and $j$ the angle of precision is not random (it is the same for all hops between these two sites). Therefore the spin orbit interaction
can be described with rotation matrixes $D_{\alpha\beta}^{ij}$. If the electron on site $i$ has
momentum expectations $M_{\alpha}$ (where index $\alpha$ stands for the cartesian coordinates),
then after the tunneling to site $j$ the expectation value of the magnetic moment is
$\sum_{\beta}D_{\alpha\beta}^{ij}M_{\beta}$. This rotation matrix should be defined for every
pair of sites  $D_{\alpha\beta}^{ij} = (D_{\alpha\beta}^{ji})^{-1}$.

To include rotation matrixes into kinetic equations (\ref{kin_f0}-\ref{kin_f2}), one should make a substitution
\begin{equation}\label{SO}
M_{\alpha}^{(j)} \rightarrow D_{\alpha\beta}^{ji} M_{\beta}^{(j)}
\end{equation}
in all kinetic equations Eqs.(\ref{kin_f0}-\ref{kin_f2}). On the other hand projections $M_{\alpha}^{(i)}$ should be unchanged.

\subsection{Generalized resistance network}
\label{sect_kinlin}

When the linear response of a hopping system to a small applied dc voltage is considered the kinetic equations can
be reduced to a resistor network. In this section we show how this network is generalized when the electron spin and the Hubbard energy are taken into account.

The reduction to the resistor network depends on the possibility to introduce quantities that change slowly in space. Note that probabilities $f_i$ and magnetic moments $M_i$ differ significantly
from site to site even in the equilibrium in the case of the broad distribution of random energies.

In the case when the directions of magnetizations on all the sites are the same \cite{comment-magn} one can introduce chemical
potentials for spin up and spin down electrons $\mu_+$ and $\mu_-$. The occupation probabilities should
be expressed as functions of these chemical potentials.
\begin{equation}\label{fmu}
 f_0^{(i)} = \frac{1}{Z_i}, \quad \rho_{+}^{+}(i) = \frac{\exp\frac{-\epsilon_i + \mu_+^{(i)}}{k_B T}}{Z_i}, \quad \rho_{-}^{-}(i) = \frac{\exp\frac{-\epsilon_i + \mu_-^{(i)}}{k_B T}}{Z}, \quad f_2^{(i)} = \frac{\exp\frac{-2\epsilon_i - U_h + \mu_+^{(i)} + \mu_-^{(i)}}{k_B T}}{Z_i}.
\end{equation}
where $Z_i$ is the statistical sum on site $i$
\begin{equation}\label{Z}
 Z_i = 1 + \exp \frac{-\epsilon_i + \mu_+^{(i)}}{k_B T} + \exp \frac{-\epsilon_i + \mu_-^{(i)}}{k_B T} + \exp \frac{-2\epsilon_i - U_h + \mu_+^{(i)} + \mu_-^{(i)}}{k_B T}.
\end{equation}
Here we assume that the magnetization of all sites is directed along $z$ axis,
therefore $\rho_{+}^{-(i)} = \rho_{-}^{+(i)} = 0$. $f^{(i)}$ and $M_z^{(i)}$ are expressed
in terms of $\rho_{+}^{+(i)}$ and $\quad \rho_{-}^{-(i)}$ according to Eq.(\ref{5bryk}).

It is easy to check by the direct substitution that when chemical potentials are
the same in all sites $\mu_+=\mu_-=\mu$ the system is in the equilibrium for any on-site
random energies $\varepsilon_i$ and all time derivatives in the kinetic equations are zero.

It is useful to have expressions for the spin up and the spin down currents. The spin up
current between sites $i$ and $j$ is given by the formula:
\begin{equation}\label{J+}
J_{ij}^+ = -e (\Gamma_{ij}^+ - \Gamma_{ji}^+),
\end{equation}
Here $\Gamma_{ij}^+$ is the hopping rate  for the spin up electrons from site $i$ to the site $j$.
\begin{equation} \label{Gamma+}
\Gamma_{ij}^+ = W_{ij} f_+^{(i)} f_0^{(j)} + W_{ij}^{+U} f_2^{(i)} f_0^{(j)} + W_{ij} f_2^{(i)}f_-^{(j)} + W_{ij}^{-U} f_+^{(i)} f_-^{(j)}.
\end{equation}
To get this expression from the kinetic equations one should consider the time derivative of the
probability $P_i^+$ to have an electron with the spin up on site $i$. Note that a double occupied
site contains electron with spin up and thus $P_i^+ = \rho_+^{+(i)} + f_2^{(i)}$. Direct
calculation yields:
\begin{equation} \label{charge-conserv}
-e\frac{d P_i^+}{dt} = \sum_{j} J_{ji}^+
\end{equation}
with Eq.(\ref{J+}) for the currents $J_{ji}^+$. The expression for spin down current can
be obtained from (\ref{J+}) and (\ref{Gamma+}) by replacing index $+$ with $-$.

The current appears when the system is placed in the electric field or when there is a shift
of the chemical potential. Expanding the kinetic equation over the small electrostatic potential
$\Delta\varphi_i$ and the small shift of the chemical potentials $\Delta\mu_i^{\pm}$ we obtain:
\begin{equation}\label{Rnet}
J_{ij}^+ = R_{ij,+}^{-1} (\Delta\varphi_{ij} - \Delta \mu_{ij}^+ /e )
\end{equation}
where $\Delta \mu_{ij}^+ = \mu_i^+ - \mu_j^+$, $\Delta \varphi_{ij} = \varphi_i - \varphi_j$
and $R_{ij,+}^{-1}$ is the effective resistance of the pair of sites $ij$ with respect to the current of the electrons with spin up. It contains four
contributions.
\begin{equation}\label{Res}
R_{ij,+} =  \left[\frac{k_B T}{e^2 \Gamma_{ij,AA}^{+,(0)}} + \frac{k_B T}{e^2 \Gamma_{ij,AB}^{+,(0)}} + \frac{k_B T}{e^2 \Gamma_{ij,BA}^{+,(0)}} + \frac{k_B T}{e^2 \Gamma_{ij,BB}^{+,(0)}}  \right]^{-1}
\end{equation}
where $\Gamma_{ij}^+$ with additional indexes $A$ and $B$ are the contributions to $\Gamma_{ij}^+$
$$
\Gamma_{ij,AA}^+ = W_{ij} f_+^i f_0^j, \quad \Gamma_{ij,AB}^+ =  W_{ij}^{-U} f_+^i f_-^j, \quad \Gamma_{ij,BA}^+ = W_{ij}^{+U} f_2^i f_0^j
$$
$$
\Gamma_{ij,BB} = W_{ij} f_2^if_-^j.
$$
The additional upper index $(0)$ in (\ref{Res}) means that values $\Gamma$ are calculated for $\Delta\varphi_{ij} =0$ and $\Delta\mu_{ij} = 0$.

The physical meaning of different contributions to $\Gamma$ becomes apparent when one considers large Hubbard energy $U \gg k_B T$. Note that in order to contribute to hopping conductivity a site should have the energy level close to the chemical potential. In this case the sites that take part in conductivity are divided into two groups \cite{Matveev}: A-sites that have $\varepsilon_i \sim \mu$ and B-sites with $\varepsilon_i + U \sim \mu$. A-sites are practically never double-occupied so for these sites $f_2 \approx 0$. For B-sites $f_0 \approx 0$, these sites never have zero electrons. For this model only one contribution to $R_{ij,+}$ is important for each pair of sites. For example if the site $i$ is of type A and the site $j$ is of type B, the resistance $R_{ij,+}$ for this pair is
$$
R_{ij,+} \approx \frac{k_BT}{e^2 \Gamma_{ij,AB}^{+,(0)}}.
$$
This result agrees with \cite{Matveev} when the temperature is much less than the Hubbard energy. We generalize this result to the case of arbitrary relation between $kT$ and $U$ and make it explicitly applicable to the case $U=0$ discussed in \cite{HF1,HF2,HF3}. We show that if the temperature is larger or comparable with the Hubbard energy each site plays both roles: of A-type and B-type. The corresponding resistances are connected in parallel.

When magnetization is not restricted to one axis and the magnetization of sites in different parts of the sample is aligned along different axis, the introduction of slowly changing chemical potentials is possible only when this axis slowly changes in space. In this case the chemical potentials $\mu_i^+$ and $\mu_i^-$ should be related to the mean spin projections on the local axis. In this case one can use the same expressions for resistors, but should keep in mind that the system kinetics cannot be reduced only to the resistor network. The situation when different parts of the sample have different magnetization axis is unstable and leads to the spin relaxation.

Finally let us note that although magnetization ${\bf M}_i$ depends (even in the equilibrium) on random energies $\epsilon_i$, the relative magnetization ${\cal M}_i = { M}_i/f_i$  is the function of the chemical potentials only.
\begin{equation} \label{m}
{\cal M}_i = \tanh\left( \frac{\mu_i^+ - \mu_i^-}{k_B T} \right).
\end{equation}

\subsection{Magnetoresistance}

A new mechanism of singlet magnetoresistance was recently proposed in organic
semiconductors\cite{HF1,HF2,HF3}. This mechanism is based on the fact that two electrons
with the same spin cannot occupy the same site even in the case of small Hubbard energy.
In Ref.\cite{HF1,HF2,HF3} it was stated that when the spin relaxation time is longer than the
hopping time, in order to hop the electron should find the site which is either free or have
an electron with opposite spin direction.
This make the hops longer than in the case when all the sites are allowed for the hop. The finite spin-flip time allows the hop to the site with the same spin projection as the hopping electron with some probability $p(H)$. The probability $p(H)$ depends on magnetic field and changes the concentration of sites included in the percolation. The concentration of sites is connected to the critical hopping length. The conductivity depends exponentially on this distance. Therefore the dependence of the probability $p(H)$ on magnetic field leads to the exponentially-strong magnetoresistance.

 The model used in Ref.\cite{HF1,HF2,HF3} does not include Coulomb interaction and any of the higher order
terms (intersite exchange interaction or under-barrier scattering). Also the importance of non-equilibrium correlated
filling numbers was not mentioned in \cite{HF1,HF2,HF3}. Therefore the kinetic equations
Eqs.(\ref{kin_f0}-\ref{kin_f2}) should be applicable to this case. We have shown that the correct
procedure is to consider averaged on-site density matrix and only then calculate the
parameters of percolation like the characteristic hopping length. As a result the magnetic
field can affect the effective resistances only via magnetization $M_i$ (at least for the case of small dc current when resistor network approximation
is applicable). When Zeeman energy can be neglected in comparison with temperature, the on-site magnetization is absent ${\bf M}_i = 0$. In this case
the spin relaxation time does not contribute to the expression for the resistances and does not
have any effect on the resulting conductivity.

The magnetoresistance appear only when Zeeman energy becomes comparable with the temperature.
In this case the stationary state have finite magnetization on each site aligned along the
magnetic field. This on-site magnetization influences the resistor network in accordance
with equation (\ref{Res}) and leads to the magnetoresistance.
However it is not a novel effect (at least for large Hubbard energy $U$). It is well known from semiconductor physics and was first reported in \cite{Kamimura}.  It is positive and a linear function of the magnetic field for $k_BT<E_Z<\xi_c k_BT$, where $\xi_c$ is the critical exponent of hopping conductivity and $E_Z=\mu_B g H$ is the Zeeman energy \cite{Matveev}. At higher fields $E_Z > \xi_c k_BT$ this magnetoresistance saturates. In the limit of the small magnetic fields $E_Z<k_BT$ it becomes quadratic $\propto (E_Z/k_BT)^2$ \cite{MyMR}.

We want to note however that our treatment is directly applicable only to the systems that are close to equilibrium. The resistor network
explicitly assumes expansion over small applied voltages. Also the Hartree decoupling applied to get the kinetic equation can be strictly justified
only near equilibrium. There is a numerical Monte-Carlo simulation \cite{Bobbert} showing that $p(H)$ can in principle influence the dc conductivity. It is important that the simulation \cite{Bobbert} deals with strongly non-equilibrium systems with the voltage applied to a single resistor at least three times larger than temperature. We believe that in order to describe the results of \cite{Bobbert} with analytical theory one should directly include the parameter of the non-linearity into considerations, because spin-blocking magnetoresistance discussed \cite{Bobbert} does not appear for small applied voltages.

We want to compare this result with the results of recent publication \cite{Spivak-new} considering magnetoresistance due to the spin precession in the hyperfine fields for another system (the model in \cite{Spivak-new} does not allow the double-occupation but include the interference of different tunneling pathes). It is shown that the effect of spin precession on the d.c. conductivity is related to the correlations of site filling numbers that appear in the non-Ohmic regime. It is out of the Hartree approximation and does not appear for small voltages.

Note that we do not provide the explanation of organic magnetoresistance within our theory. We demonstrate that in the limit of the small electric field where the reduction of the kinetic equations to the effective resistor network is justified the magnetoresistance appears only when magnetization is finite. In order to describe the magnetoresistance \cite{Bobbert} that appears without average magnetization it is necessary to go to the limit of  strong electric field where the site occupation numbers may be correlated and our kinetic equations are not justified.

\section{Spin relaxation due to the random fields}
\label{sect_relax}

In order to demonstrate that the kinetic equations Eqs.(\ref{kin_f0}-\ref{kin_f2}) are a useful tool
to understand the physics of hopping conduction we apply them to the problem of spin relaxation
in disordered semiconductors with the hopping transport.

We consider the
simple case of neighbor hopping. It means that we assume the temperature to be larger than the width
of the distribution of site energies $\varepsilon_i$. In this case the energy disorder can be neglected and
the positional disorder define the distribution of hopping rates. We show that even in this simple case there is a number
of novel effects that were not discussed previously. This effects are related to the exponentially
broad distribution of the hopping rates.

Numerical studies of the hopping conduction usually consider a set of sites with random energies on
a lattice. Thus the spatial disorder is totaly ignored and only energy disorder is considered.  However at least one case is known when  the spatial and the energy disorder lead to different results. It is the case of slow relaxation in systems with the hopping conduction and the strong Coulomb interaction \cite{Efr-Ts}.  It gives us additional reason to focus on the positional disorder in the present paper.

We consider a set of identical sites that are randomly distributed in space with the hopping
probability exponentially decaying with the distance between sites $r_{ij}$,
$W_{ij} \propto W_0\exp(-2r_{ij}/a)$, where $a$ is the localization radius. At each site there
is a random hyperfine magnetic field $\vec{\cal H}_i$ with the characteristic scale $\langle {\cal H} \rangle$.
Initially the system is in the equilibrium. At $t=0$ all sites obtain small
magnetization ${\bf M}_i$ aligned along z-axis. As far as it is small the relaxation of magnetic
moments does not influence occupation probabilities $f_0$, $f_1$, $f_2$. The system is defined
by two parameters:  the conventional localization  parameter $na^3$ and the ratio of the
characteristic precession frequency in the hyperfine magnetic field to the hopping rates
$\langle {\cal H} \rangle/W_0$. In the case of the variable range hopping conductivity (when the energy disorder is essential) one should
also keep in mind the relation between the characteristic size of random energy, Hubbard energy and the temperature. These parameters however are out of the scope of our work.

Let us write the general equation for the spin relaxation
\begin{equation} \label{Spin-rel-gen}
\frac{d M_\alpha^{(i)}}{dt} + \epsilon_{\alpha\beta\gamma}M_\beta^{(i)} {\cal H}^{(i)}_\gamma = \sum_{j} \Upsilon_{ji} M_\alpha^{(j)} - \Upsilon_{ij} M_\alpha^{(i)},
\end{equation}
\begin{equation}\label{Upsilon}
\Upsilon_{ij} = W_{ij}f_0^{(j)} + W_{ji}f_2^{(j)} + \frac{W_{ij}^{-U}}{2}f_1^{(j)} + \frac{W_{ji}^{-U}}{2}f_1^{(j)}.
\end{equation}
Here $\Upsilon_{ij}$ is the rate of spin transition from site $i$ to site $j$. Note that in
general case $\Upsilon_{ij} \ne \Upsilon_{ji}$ even in equilibrium. It is related to the fact
that different sites have different equilibrium probability $f_1$ to be single-occupied and
thus different equilibrium magnetization. However in the considered problem the sites are
equivalent and therefore the spin transfer rate is directly connected with the charge
transfer rate
$$
\Upsilon_{ij} = \Upsilon_{ji} \propto R_{ij}^{(-1)}.
$$

\subsection{Spin relaxation in a pair of sites.}

The elementary source of the magnetic moment relaxation is a pair of sites with different
local hyperfine magnetic fields. These sites are connected by the spin transition rate
$\Upsilon$. The magnetization dynamics of these sites is described by the equations:
\begin{equation} \label{pair}
\frac{d {\bf M}_1}{dt} + \vec{\cal H}_1 \times {\bf M}_1 = \Upsilon({\bf M}_2 - {\bf M}_1),
\end{equation}
$$
\frac{d {\bf M}_2}{dt} + \vec{\cal  H}_2 \times {\bf M}_2 = \Upsilon({\bf M}_1 - {\bf M}_2).
$$
Let us discuss the relaxation in the two limiting cases: when the precession frequency is larger
than the tunneling rate ${\cal H}_{1,2} \gg \Upsilon$ and in the opposite limit ${\cal H}_{1,2} \ll \Upsilon$.

In the case of the strong magnetic field the moments precess around the
local fields. The relaxation of
their absolute values is governed by the tunneling rate $\Upsilon$. Let us assume that at $t=0$,
$|{\bf M}_1| = |{\bf M}_2| = M$. It is easy to show that relation $|{\bf M}_1| = |{\bf M}_2|$
holds during the relaxation. Therefore we may write only the equation for $M$
\begin{equation}\label{pairH}
\frac{d M}{d t} = - \Upsilon (1-\cos(\angle M_1 M_2)) M.
\end{equation}
Here $\angle M_1 M_2$ is the angle between magnetic moments ${\bf M}_1$ and ${\bf M}_2$. In the
case of fast precession it should be averaged over precession period. Assuming that
$|{\cal H}_1| \ne |{\cal H}_2|$ yields:
\begin{equation}\label{PairH1}
\frac{d M}{d t} = - \Upsilon \left(1- \frac{M_{1\|}}{M_1} \frac{M_{2\|}}{M_2}\cos(\angle {\cal H}_1 {\cal H}_2) \right) M.
\end{equation}
Here $M_{1\|}(M_{2\|})$ is the component of magnetic moment ${\bf M}_1({\bf M}_2)$ aligned along the local field
$\vec{\cal H}_1(\vec{\cal H}_2)$.

The relaxation rate is zero for the case when both magnetic field and magnetic moments are
aligned along the same axis and ${\bf M}_1 = {\bf M}_2$. In all other cases it is finite
and proportional to $\Upsilon$.

In the opposite case $\Upsilon \gg {\cal H}$ in the initial relaxation phase $t \sim 1/\Upsilon$ the
difference of magnetic moments ${\bf M}_1 - {\bf M}_2$ relaxes with the fast relaxation rate
$\Upsilon$. After this phase of relaxation the moments are different only because of finite local
fields and  ${\bf M}_1 - {\bf M}_2 \propto {\cal H}/\Upsilon$. However the average magnetic moment
${\bf M}_{+} = ({\bf M}_1 + {\bf M}_2)/2$ cannot relax with the rate $\Upsilon$. The equation
for ${\bf M}_{+}$ up to the terms $\propto {\cal H}^2/\Upsilon$ has the following form:
\begin{equation} \label{pairG}
\frac{d \bf{M}_+}{dt} + \vec{\cal H}_+ \times {\bf M}_+   - \frac{1}{2\Upsilon} [\vec{\cal H}_- \times [\vec{\cal H}_- \times {\bf M}_+]]= 0.
\end{equation}
Here $\vec{\cal H}_{\pm} = (\vec{\cal H}_1 \pm \vec{\cal H}_2)/2$.

The magnetic moment ${\bf M}_{+}$ precess around average magnetic field, as it can be seen from
the second term of the equation. The third term gives a small alternation to this precession,
but more importantly, it yields relaxation of the absolute value $M_+ = |{\bf M}_+|$.
$$
\frac{d M_+}{dt} = - \frac{{\cal H}_-^2}{2\Upsilon} \left(1 - \frac{(\vec{\cal H}_- \cdot\vec{\cal H}_+)^2}{{\cal H}_+^2 {\cal H}_-^2} \frac{(\vec{\cal H}_+ \cdot{\bf M}_+)^2}{{\cal H}_+^2 M_+^2} \right)M_+
$$
The relaxation of the magnetic moment is proportional to ${\cal H}^2/\Upsilon$. It is the motion
suppression of relaxation well known from \cite{motional_narrowing}, where it is related to the
electron diffusion. Here we show that for considered problem the diffusion over macroscopic distances is not required
for this suppression. It appears even when tunneling between two sites is considered.

\subsection{System without disorder}

\begin{figure}[htbp]
    \centering
        \includegraphics[width=0.7\textwidth]{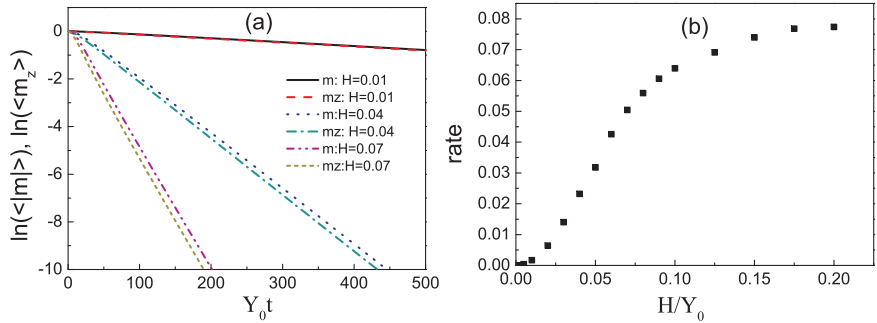}
        \caption{The relaxation of the magnetic moment of a hopping system without disorder. (a) The dependence of logarithm of $\langle |m| \rangle$ and $\langle m_z \rangle$ on time for different values of random magnetic field. (b) The dependence of relaxation rate on average on-site magnetic field.  }
    \label{fig:lat}
\end{figure}

A system with a large number of sites with low disorder can be characterized by some average
spin transition rate $\overline{\Upsilon}$. The spin relaxation in such a system is analogous
to the spin relaxation in a pair of sites. For the case of fast hopping
$\overline{\Upsilon} \gg {\cal H}$ the magnetization is aligned along the initial magnetization
axis and slowly precess around average magnetic field (that tends to zero for a macroscopic
system). As a results the magnetization relaxes with the rate $\propto {\cal H}^2/\Upsilon$.

For a large magnetic field slow tunneling cannot keep site moments out of the precession around their local fields. Due to this precession the macroscopic moment of the system is decreased by a factor of 3, according to Kubo-Toyabe formula \cite{KuboToyabe}. On the other hand the local magnetic moments on sites remains. The following relaxation of the magnetic moment goes with the rate $\Upsilon$ that does not depend on the magnetic field.

In Fig. (\ref{fig:lat}) the results of numerical solution of Eq. (\ref{Spin-rel-gen}) for the cubic lattice are shown. The considered system has the following set of parameters $\Upsilon_0 = 1$, $na^{1/3} = 0.5$ and $n = 1$. The spin transition rate between neighboring sites is $\Upsilon_{neib} = 0.018$. In the model we use linearized kinetic equation considering the initial magnetization to be small $M_i(0)\ll 1$. In Fig. (\ref{fig:lat}) we plot relative magnetizations ${\bf m}_i(t) = {\bf M}_i(t)/|{\bf M}_i(0)|$. Naturally at the beginning of the simulation for every site $|{\bf m}_i| =1$ and all ${\bf m}_i$ are aligned along z-axis. In the linear case ${\bf m}_i$ follow the same equations as ${\bf M}_i$.

We track independently averaged absolute value of the on-site magnetic moment $\langle |m| \rangle$, and averaged z-component of the magnetic moments (that correspond to the macroscopic magnetization of the system). It can be seen that during the first phase of the relaxation $\langle |m|\rangle$ and $\langle m_z\rangle$ become slightly different due to random precession in magnetic field. However random magnetic field does not reduce the ratio $\langle m_z\rangle/\langle |m|\rangle$ below $1/3$. The hopping makes this ratio even larger. Then the relaxation of $\langle m_z\rangle$ follows the relaxation of absolute values of the on-site moments. The relaxation has exponential form with some relaxation rate.

In Fig.(\ref{fig:lat}b) we show the dependence of relaxation rate on average magnetic field. The dependence is quadratic for small fields and saturates when oscillation in random fields becomes much faster than the spin transition. This result agrees with the two regimes of slow and fast hopping described in \cite{Yu-hyperfine}.

\begin{figure}[htbp]
    \centering
        \includegraphics[width=0.4\textwidth]{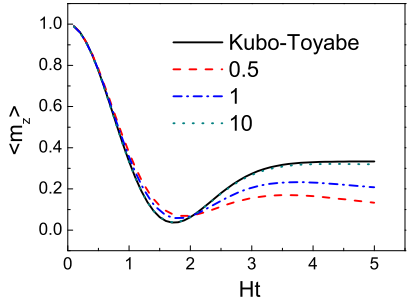}
        \caption{The relaxation of the magnetic moment of a hopping system without disorder for large random magnetic compared with Kubo-Toyabe formula. The numbers correspond to the parameter $H/\Upsilon_0$.   }
    \label{fig:kubo}
\end{figure}

At very large magnetic fields a minimum appears in the time dependence of $\langle m_z \rangle$ in agreement with \cite{KuboToyabe}. On Fig. \ref{fig:kubo} the initial part of relaxation of $\langle m_z \rangle$ is shown. When the precession in the magnetic field is much faster than the hopping it is more instructive to plot relaxation not versus $\Upsilon_0 t$ but versus $Ht$. For very strong magnetic field (or very slow hops) the computations agree with Kubo-Toyabe formula (Fig. \ref{fig:kubo}). The finite hopping rate smears this dependence. We want to note that the relation of precession frequency to the hopping rate corresponding to the appearance of minimum in $\langle m_z \rangle$ is really large. In the discussed results neighbor spin transition rate is $\approx 0.018\Upsilon_0$ and the relation $H/\Upsilon_0 = 0.5$ correspond to the precession frequency that is $\sim 30$ times larger than the spin transition rate.

\subsection{Random positions of sites}

The system with the positional disorder is characterized by the parameter $na^3$ where $n$ is the site concentration and $a$
is the localization length. For large $na^3 \sim 1$ the hopping rates to the neighbors are of the
same order for all sites. In this case the disorder becomes relatively unimportant. Without
polaronic effects this system exhibit the metal-insulator transition and its conductivity is of
the band type. Due to polaron formation the transport mechanism in a system with low disorder can
still be of the hopping type but most interesting part of the physics that is usually associated
with hopping conduction is absent. Therefore we expect that the system should behave
similarly to the hopping systems without disorder discussed in previous section.

For the small values of $na^3$, when the characteristic distance to the nearest neighbor is larger than
the localization length $r_{nn} \sim n^{-1/3} \gg a$, the hopping rates (even without energy
disorder) have exponentially broad distribution. Transport in that case is usually described in terms of the percolation theory. The conduction is governed by the
threshold hopping rate $\Gamma_{perk}$ that allows the percolation over the macroscopic distances.
Most of the current is carried by the infinite cluster of the sites that are connected with
hopping rates $\Gamma \gtrsim \Gamma_{perk}$. The density of the infinite cluster tends to zero for
$na^3 \rightarrow 0$. This small portion of sites that form the infinite cluster plays important
role in the theory of variable range hopping conductivity. However for the problem of neighbor hopping the infinite cluster density becomes small only for very small localization length $n^{1/3}a < 0.1$ \cite{PikeSeager}.

For the systems with low disorder there are two possible relation of parameters
$\langle {\cal H} \rangle < \overline{\Upsilon}$ and $\langle {\cal H} \rangle>\overline{\Upsilon}$. The first relation leads
to the motion suppression of the relaxation with relaxation rate
$\sim \langle {\cal H} \rangle^2/\overline{\Upsilon}$. The second relation leads to the relaxation rate
$\sim \overline{\Upsilon}$.  When the distribution of the hopping rates $\Gamma_{ij}$ and the
spin transfer
rates $\Upsilon_{ij}$ is exponentially broad, the natural situation is that this inequality
have different sign for different hops.

In this case one can assume that in terms of the spin relaxation the sites should be arranged
into clusters. Inside the cluster the spin transfer rates $\Upsilon > \langle{\cal H}\rangle$ and the magnetic moments
in the cluster have approximately the same direction. Different clusters are connected by a
relatively slow hopping rates $\Upsilon<\langle {\cal H} \rangle$. Thus the relaxation will be governed by the critical
transfer rates $\Upsilon_{th} \sim \langle {\cal H} \rangle$. The relaxation due to the fast transfer rates
$\Upsilon \gg \langle {\cal H} \rangle$ is suppressed by the motion and transfer rates that are smaller
than precession rate $\Upsilon \ll \langle {\cal H} \rangle$ are too slow to influence the relaxation significantly.
Therefore it is natural to assume that the dependence of spin relaxation on the characteristic
scale of random magnetic field should be $\propto \langle {\cal H} \rangle$, or may be
$\propto \langle {\cal H} \rangle^\alpha$, where $\alpha < 2$.
Note that a statement that spin relaxations is governed by the pairs of sites with the hopping rate comparable to the spin precession frequency appears in \cite{Lyubinskiy} although only for the pairs of sites that are well separated from the rest of system.

In order to study spin relaxation we perform extensive numerical calculations based on the
general kinetic equation (\ref{Spin-rel-gen}). We consider numerical samples with random uncorrelated positions of
sites. Each site is ascribed by a random local magnetic field. The spin transition rates between
sites are exponentially decaying with distance
$$
\Upsilon_{ij} = \Upsilon_0 \exp(-2r_{ij}/a).
$$
We use cutoff at some large distance $r_{max}$. $r_{max}$ is chosen in such way that each site
is connected in average with 20 other sites. This choice of $r_{max}$ allows us to consider
relatively large systems up to $10^4$ sites. More details about numerical computations are given in appendix.

We start from the case when all the magnetic moments are aligned
along $z$ axis and consider the relaxation of the relative magnetizations ${\bf m}_i(t) = {\bf M}_i(t)/|{\bf M}_i(0)|$ in time. We consider three
different values of parameter $n^{1/3}a$: $0.2$, $0.5$ and $1$. For $n^{1/3}a = 0.2$ ($na^3 = 8\cdot10^{-3}$) the
system is deeply in the localized regime even without formation of polarons. Percolation hopping
rate $\Gamma_{perk}$ is four orders of magnitude less than $\Gamma_0$. However even for this
small $na^3$ the density of infinite cluster is rather high $\sim  0.85$, yet the
amount of sites that are out of the infinite cluster is significant. For the value $n^{1/3}a = 0.5$
($na^3 = 0.125$) the hopping distances are still larger than the localization length and
exponentially broad distribution of hopping rates persist. However the density of infinite
cluster is $>0.95$ and there is only small amount of sites that are out of this cluster. Finally
for $n^{1/3}a=1$ there is no exponential distribution of neighbor hopping rates. In this case
hopping transport is possible only due to formation of polarons.

\begin{figure}[htbp]
    \centering
        \includegraphics[width=0.9\textwidth]{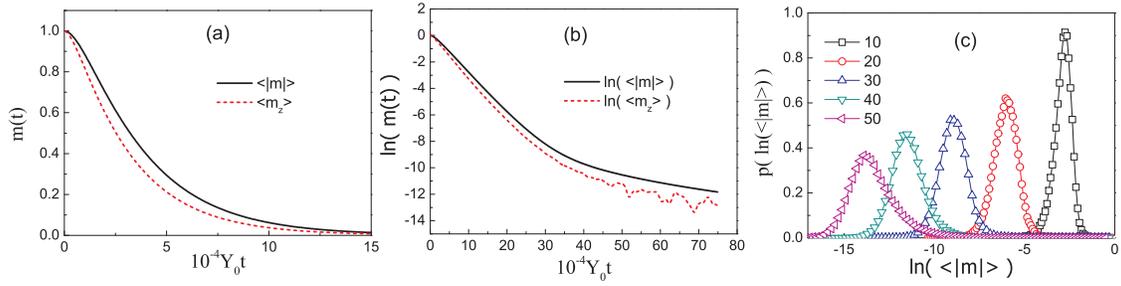}
        \caption{The relaxation of magnetic moment for $n^{1/3}a = 0.2$ and  $\langle{\cal H}\rangle = 10^{-4}\Upsilon_0$. (a) the values of average absolute value of site magnetic moment $\langle |m|\rangle$ and average $z$-projection of magnetic moment $\langle m_z\rangle$ at the initial part of the relaxation. (b) the logarithms of $\langle |m|\rangle$ and $\langle m_z\rangle$ for all the computed time. (c) the distribution of absolute values of magnetic moments at time $\Upsilon_0t \cdot 10^{-4} = $ $10$, $20$, $30$ and $40$.   }
    \label{fig:1}
\end{figure}

For the small localization length $n^{1/3}a = 0.2$ the percolative value of the spin transition
rate $\Upsilon_{perk} \approx 1.8\cdot 10^{-4}\Upsilon_0$. We start the discussion from the case
when the characteristic size of local hyperfine magnetic fields is slightly below of
$\Upsilon_{perk}$, $\langle{\cal H}\rangle = 10^{-4}\Upsilon_0$.  The results of calculations are shown
in Fig. \ref{fig:1}. We averaged the results over 50 numerical samples with different random
positions of sites (each sample contains $10^4$ sites).

At the initial phase of the relaxation $t \lesssim  {\cal H}^{-1}$, there is little change in
$\langle |m| \rangle$ while $\langle m_z \rangle$ decreases due to rotation of the magnetic
moments in the hyperfine magnetic field. However, the rotation in the random magnetic field alone
cannot decrease $\langle m_z \rangle$ more that to $1/3$. Therefore subsequent
relaxation of $\langle m_z \rangle$ accompanies the relaxation of $\langle |m| \rangle$.

Then for a large time interval ${\cal H}t<20$, the relaxation is exponential. However at larger
time the exponential relaxation slows down. To understand
this behavior we computed the distribution of absolute values of on-site magnetization at
different $\Upsilon_0t$  (Fig. \ref{fig:1} (c)). This distribution becomes exponentially broad
for $\Upsilon_0 t>20$ and the average magnetic moment is determined by a small number of sites with the
largest magnetic moment. This behavior is quite natural for a system with the
exponential distribution of the hopping times.

In Ref.\cite{ShklovskiiTraps} Shklovskii has introduced a concept of "traps" that appear in the
hopping systems. The typical trap is a pore with relatively large radius that can randomly appear in a random site distribution. There is
one site inside the pore that act as a trap. The capture/relese rate of the trap is related to the hopping rate from
the site in the pore to the outside sites. In
Ref.\cite{ShklovskiiTraps} this concept was applied to explain $1/f$ noise in the hopping
systems. We argue that the same traps can be responsible for the spin relaxation at large
times. Naturally, the spin on these traps cannot relax faster than $\Upsilon_{trap}$, where
$\Upsilon_{trap}$ is the highest spin transfer rate from the trap to the surroundings. In section \ref{sect_aprox} we show
that the spin relaxation due to these traps is non-exponential and derive the analytical approximation for this non-exponential
relaxation.

Note that the initial exponential part of the relaxation appears to be self-averaged even for one numerical sample consisting of $10^4$ sites. However for larger times it becomes more difficult  to obtain averaged results. For $\Upsilon_0 t > 50 \cdot 10^4$ we observed oscillation in $\langle m_z \rangle$ that are due to the lack of disorder averaging. We believe that this slow suppression of these oscillations with
averaging over disorder realizations is related to the trap-dependent relaxation. The characteristic size of the trap that governs magnetic moment at time $t$ grows with time. However the probability to find corresponding trap exponentially decreases with the trap size leading to the poor averaging of magnetic moment at large times.

\begin{figure}[htbp]
    \centering
        \includegraphics[width=0.6\textwidth]{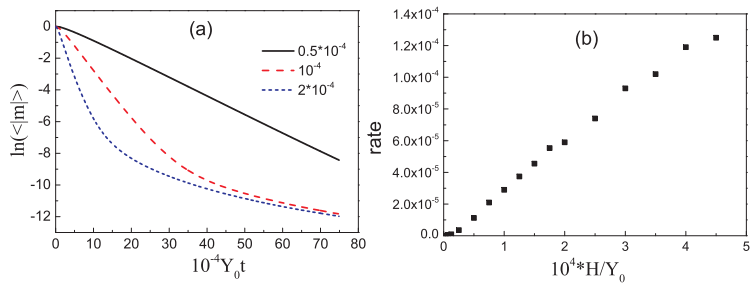}
        \caption{The relaxation of the magnetic moment for $n^{1/3}a=0.2$ and different magnetic fields $\langle{\cal H}\rangle/\Upsilon_0$ (a) and the dependence of the relaxation rate (in the exponential phase of relaxation) on the magnetic field  (b). }
    \label{fig:mag}
\end{figure}
Let us now discuss the dependence of the relaxation on the characteristic value of the magnetic field. This dependence is shown in Fig. \ref{fig:mag}. It is seen from this figure that the final phase of the relaxation at large $t$ is essentially independent of the magnetic field. The rate of the exponential part of the relaxation depends linearly on the magnetic field,  as it was predicted above on the basis of simplified arguments. The characteristic time of the transition from exponential to non-exponential relaxation decreases with increasing field. For large magnetic field the relaxation is non-exponential at any time.

\begin{figure}[htbp]
    \centering
        \includegraphics[width=0.6\textwidth]{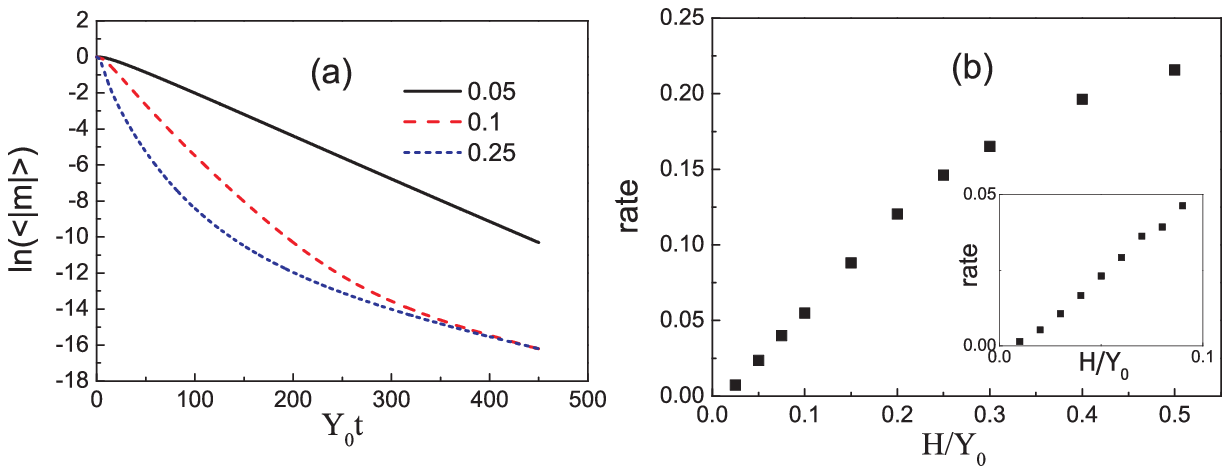}
        \caption { (a) Spin relaxation for $n^{1/3}a=0.5$ for different magnetic fields. Numbers on plots correspond to $\langle {\cal H} \rangle/\Upsilon_0$. (b) The dependence of the relaxation rate (in the exponential regime) on the characteristic scale of the random field. }
    \label{fig:a05}
\end{figure}

Consider now the other values of the parameter $n^{1/3}a$. The results
of computation of spin relaxation for intermediate localization length $n^{1/3}a = 0.5$  are shown in Fig. \ref{fig:a05}. The qualitative
picture of the relaxation is similar to the case of small localization length $n^{1/3}a = 0.2$.
The relaxation is exponential for the short times and then follows universal non-exponential curve independent from the value of random field. The rate of the exponential relaxation is proportional to the magnetic field for a wide range of fields
$0.01<\langle {\cal H}\rangle/\Upsilon_0 < 0.2$ (Fig. \ref{fig:a05}(b)).  At fields $\langle {\cal H}\rangle/\Upsilon_0 > 0.2$ the linear
dependence starts to saturate.  However for these fields exponential part of the relaxation is rather small. In this part spin decrease is less than one order of magnitude. For $\langle {\cal H}\rangle/\Upsilon_0 > 0.5$ it is impossible to separate
the exponential part of the relaxation. The conductive cluster (with the above formal definition) consist of
$\sim 98\%$ of sites making the concept of percolative cluster ill-defined. Therefore we do not discuss the difference between relaxation
of sites within and outside of the percolative cluster for $n^{1/3}a = 0.5$.

\begin{figure}[htbp]
    \centering
        \includegraphics[width=0.6\textwidth]{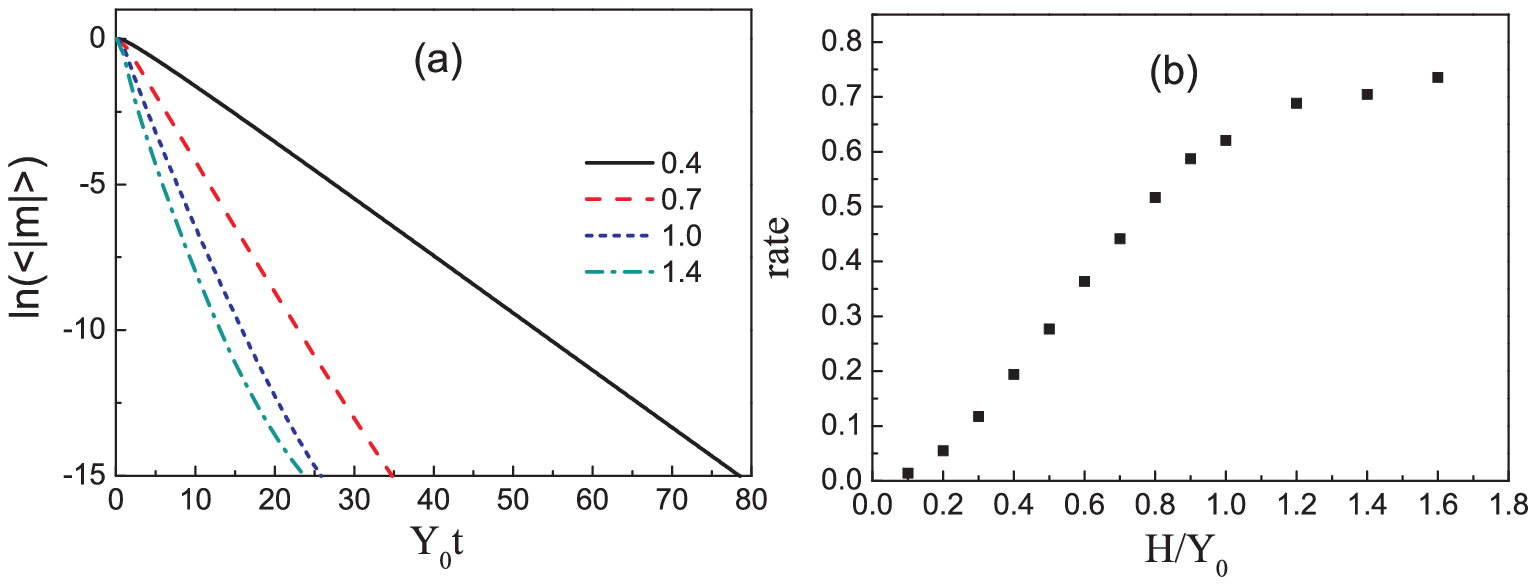}
        \caption { (a) Spin relaxation for $n^{1/3}a=1$ for different magnetic fields. Numbers on plots correspond to $\langle {\cal H} \rangle/\Upsilon_0$. (b) The dependence of the relaxation rate on the characteristic scale of the random field. }
    \label{fig:a1}
\end{figure}

Finally at fig. \ref{fig:a1} we present the results of relaxation computation for $n^{1/3}a = 1$. They  are quite similar to the relaxation for hoppnig system on the lattice. However the relaxation in the limit of strong random field is slightly non-exponential.

\subsection{Spin relaxation and the infinite percolative cluster}

We have shown that the spin relaxation in the systems with the hopping conductivity is extremely non-homogenous. Spin on different sites relaxes with different rates. Note that significance of different sites in other situations, for example for the electrical current, is also different. The current is carried by the infinite percolative cluster. Therefore it is interesting to know whether the relaxation on the sites of the infinite cluster is different from the relaxation on the sites out of this cluster. In our study we use the following formal definition of the conductive cluster (for details see \cite{Efr-Sh}). First we find the percolative threshold, i.e. the critical distance $r_{th}$ that allows the percolation across the numerical sample with hops over distances $r \le r_{th}$. Then we consider sites to be connected if the distance between them is less or equal than $r_{th} + a/2$ (it ensures that hopping rate between these sites is higher or comparable with the critical percolative rate). Finally we find the infinite cluster connected with distances $r \le r_{th} + a/2$. We consider this cluster to be the percolative cluster responsible for conduction.
The density of this cluster appears to be relatively high ($\approx 0.85$) for $n^{1/3}a = 0.2$.

\begin{figure}[htbp]
    \centering
        \includegraphics[width=0.6\textwidth]{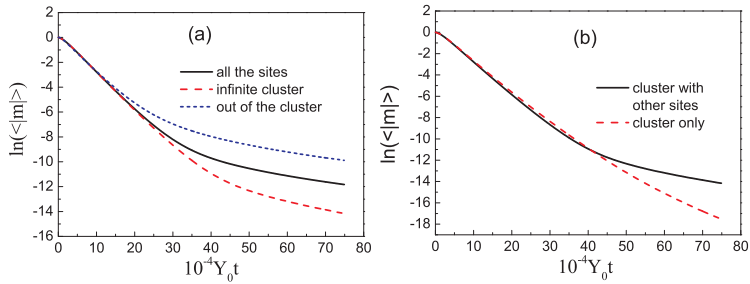}
        \caption{ (a) the relaxation of the magnetic moment inside and outside of the infinite cluster. (b) the contribution of sites out of the cluster to spin relaxation in the cluster}
    \label{fig:clu}
\end{figure}

We perform the averaging of the magnetic moments
independently for the sites of the infinite cluster and for the sites outside of the infinite cluster
and compare the results with the magnetic moment averaged over all the sites (Fig.
\ref{fig:clu} (a)). One can see that initially ($\Upsilon_0 t \le 2.5\cdot 10^3$) the relaxation
of the whole numerical sample follows the relaxation of the infinite cluster. It is natural
because most of sites belong to this cluster. However for $\Upsilon_0 t > 2.5\cdot 10^3$ these
relaxations start to deviate one from another and magnetization of the whole sample appears to be larger than magnetization
of the infinite cluster. In this time domain the
relaxation is governed by the spin relaxation on the relatively rare sites with the slow
relaxation rate that are outside of the infinite cluster. The relaxation of the average moment slows down
while the relaxation of the magnetic moment on the infinite cluster is still exponential until $\Upsilon_0 t\sim 4\cdot 10^3$.

Most interesting is that for larger times $\Upsilon_0 t > 4\cdot 10^3$, the relaxation of
sites in the infinite cluster also slows down. We believe that the reason for this slowing down
of the relaxation is the spin transfer from slow relaxing traps to the infinite cluster which
leads to the re-magnetization of the cluster. Actually, the traps can not lose magnetization by
themselves. They slowly transfer the magnetic moment to the sites of the infinite cluster where
it relaxes. To prove that we remove from the system all sites that do not belong to the infinite
cluster and recalculate the relaxation.  We compare the results with the average magnetic moment
on the cluster when all the sites in the system are present in Fig. \ref{fig:clu} (b). For times
$\Upsilon_0 t < 4\cdot 10^3$ the curves are the same except for the small difference in
relaxation time. However for $\Upsilon_0 t > 4\cdot 10^3$ the relaxation of cluster slows down
when all sites are included into the computation and remains exponential when we exclude
all sites outside of the cluster. Therefore we conclude that $\approx 15\%$ sites that are not included in the infinite
cluster govern the magnetization dynamics of the infinite cluster at sufficiently large times.

In the discussed situation the difference between relaxation of the average moment and of the moment of the infinite cluster
become important only when the magnetic moment of the system becomes small ($\sim 10^{-4}$ from the initial magnetization). We believe that
it is due to the fact that in our situation most sites are included into the percolation cluster. However the density of the cluster tends to
zero for $na^3 \rightarrow 0$. Also it is known that the density of the infinite cluster is considered to be small in the theory of variable-range hopping \cite{Efr-Sh}. We believe that when the density of the percolative cluster is small the difference between the mean magnetization and the magnetization of the infinite cluster should be more pronounced.

\subsection{Analytical approximation for non-exponential relaxation.}
\label{sect_aprox}

We have demonstrated that even at the small magnetic fields the relaxation of the magnetic moment have
slow non-exponential tails. At large values of the random magnetic fields these non-exponential
regime covers most part of the relaxation. We have argued that at least for large times this
non-exponential relaxation can be described in term of traps, the sites that are rather
far from their neighbors. Let us discuss the physics of this non-exponential relaxation.

Let us consider the spin on some trap. When the magnetic moment is transferred from the trap to
its neighbors it quickly relaxes due to fast hops outside of the trap. Therefore each trap $i$ has
the relaxation rate $\gamma_i$ that is the rate of electron hopping from this trap.

Naturally $\gamma_{i}$ is proportional to the $\exp(-2r_{neib}^{(i)}/a)$, where $r_{neib}^{(i)}$
is the distance between the trap and the nearest site. It is important that
$r_{neib}^{(i)} \gg a$. The characteristic number of sites that are effective neighbors of the
trap is $\sim  4\pi n (r_{neib}^{(i)})^2a$. For large $r_{neib}^{(i)}$ it becomes larger than
unity. In this case $\gamma_i$ can be expressed as
\begin{equation}\label{nei_gamma}
\gamma_i = \Upsilon_0\int_{r_{neib}^{(i)}}^{\infty} e^{-2r/a} 4\pi r^2 n dr = \Upsilon_0 \frac{\pi n a^3}{2}\left(2 + \frac{4r_{neib}^{(i)}}{a} + \frac{4 (r_{neib}^{(i)})^2}{a^2} \right) e^{-2r_{neib}^{(i)}/a}.
\end{equation}

\begin{figure}[htbp]
    \centering
        \includegraphics[width=0.5\textwidth]{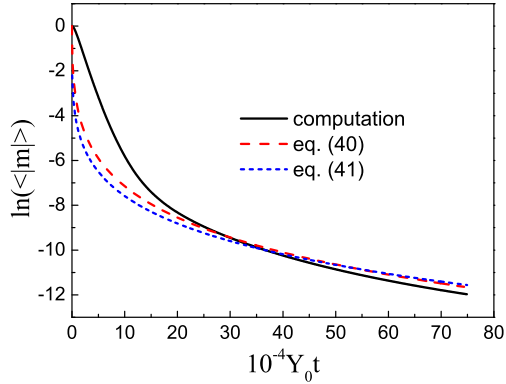}
        \caption{ Spin relaxation for $n^{1/3}a=0.2$, $\langle {\cal H}\rangle/\Upsilon_0 = 2\cdot 10^{-4}$ compared with analytical approximation (\ref{mPores}) and (\ref{nei_rate}) }
    \label{fig:model}
\end{figure}

To estimate the relaxation one should integrate the exponent $e^{-\gamma_i t}$  with the
distribution function of distances to the nearest neighbor.
\begin{equation}\label{mPores}
\langle|m(t)|\rangle = \int_0^{\infty} 4\pi nr^2 \exp\left( -\Upsilon_0 t e^{-2r/a} \frac{\pi n a^3}{2}\left(2 + \frac{4r}{a} + \frac{4 r^2}{a^2} \right) - \frac{4\pi nr^3}{3} \right)dr
\end{equation}
The integral (\ref{mPores}) can be approximated in the limit $r_{neib}^{(i)} \gg a$,
\begin{equation}\label{nei_rate}
\langle |m| \rangle(t) = m_0\exp\left( - \frac{\pi n a^3}{6} \ln^3(\Upsilon_0 t) \right) .
\end{equation}

Fig. \ref{fig:model} shows the comparison of the numerical simulations of the relaxation of the magnetic moment for $n^{1/3}a = 0.2$, $\langle {\cal H}\rangle/\Upsilon_0 = 2 \cdot 10^{-4}$ with the approximate formulae (\ref{mPores}) and (\ref{nei_rate}). The comparison with Eq.(\ref{mPores}) contain no free parameters, comparison with Eq.(\ref{nei_rate}) contain one free parameter $m_0$. We conclude that there is at least semi-quantitative agreement between the simulation in the non-exponential phase of relaxation and approximate formulae.

Eq.(\ref{nei_rate}) is in agreement with the expression proposed in Ref.\cite{Dmitriev}
for the case of the spin-orbit mechanism of relaxation in the limit of strong spin-orbit
coupling. Note that the similar expression in \cite{Dmitriev} has more free parameters.  Although our theory and the theory presented in Ref.\cite{Dmitriev} deal with different relaxation mechanisms, in the limit of strong magnetic field or the limit of strong spin orbit interaction the relaxation in both cases is governed by the distribution of hopping rates. As a result different spin relaxation mechanisms lead to the similar non-exponential regimes of the spin relaxation.

\section{Conclusion}

We derived the kinetic equations for the hopping transport that take into account the electron spin and the possibility of double occupation. In the limit of low voltage they are reduced to the generalized Miller-Abrahams resistor network. We have applied the kinetic equations to the problem of spin relaxation in the positionaly disordered system with neighbor hopping due to interaction with the random on-site hyperfine (or fringe) magnetic field. We show that the initial relaxation rate is governed by the critical hops with the rates comparable with the rate of the spin precession in a random fields. At large times as well as in the case of the large random fields the relaxation becomes non-exponential and is related to the relaxation of the spin in the traps. The relaxation is strongly inhomogeneous: the relaxation of sites in the conduction cluster differs substantially from the relaxation of the sites outside of this cluster. However in some cases the traps can drastically affect the relaxation of the magnetic moment of the infinite cluster.

\section{Appendix 1: Kinetic equation coefficients with double occupation and spin }

Here we present the derivation of matrix elements $W_{xyz}$ in the equation (\ref{kin-gen})
\begin{equation}\label{A-kin-gen}
\frac{d \rho_x^{(i)}}{d t} - {\cal S}_{xy}(i) \rho_{y}^{(i)} = \sum_j W_{xyz}(ij)\rho_y^{(i)}\rho_z^{(j)}.
\end{equation}
where $x$, $y$ and $z$ can correspond to one of $f_0$, $f_1$, $f_2$, $M_x$, $M_y$ or $M_z$.
 The straightforward calculation of matrix elements with equation (\ref{colint}) is rather cumbersome. The two-site density matrix is a $16\times16$ matrix with 256 matrix elements. Therefore we introduce a trick to find elements $W_{xyz}$.

Let us note that equation (\ref{A-kin-gen}) should be valid for any density matrices $\rho_x(i)$
and $\rho_y(j)$. However there are some special cases when this equation is reduced to a more
simple form. Consider for example tunneling from a site $i$ that is single-occupied with an
electron with spin up (at some moment $t_1$) to a site $j$ that is free at this moment. It
means that at $t=t_1$,
$$
f_1^{(i)} = 1, \quad M_z^{(i)} = 1, \quad f_0^{(j)} = 1.
$$
Other elements of $\rho(i)$ and $\rho(j)$ are equal to zero. The transition probability in this case is $W_{ij}$ described in Eq.(\ref{W}). Therefore one can write:
$$
\frac{d f_1^{(j)}}{dt} = W_{ij}, \quad \frac{d M_z^{(j)}}{dt} = W_{ij}, \quad \frac{d f_0^{(j)}}{dt} = - W_{ij}.
$$
The derivatives of other elements of $\rho^{(j)}$ are equal to zero. The same situation is described with equation (\ref{kin-gen}) at the moment $t_1$ as:
$$
\frac{d f_1^{(j)}}{dt} = W_{f_1, f_0, f}(ij) + W_{f_1, f_0, M_z}(ij), \quad \frac{d M_z^{(j)}}{dt} = W_{M_z, f_0, f_1}(ij) + W_{M_z, f_0, M_z}(ij),
$$
$$
\frac{d f_0^{(j)}}{dt} = W_{f_0, f_0, f}(ij) + W_{f_0, f_0, M_z}(ij).
$$
As a result we obtain three equations for the hopping rates $W_{xyz}(ij)$.

Considering tunneling of the electron with other spin projections from a single-occupied site to a free site we obtain a set of equations
\begin{equation} \label{e1}
\begin{array}{l}
W_{f_1,f_0,f}(ij) \pm W_{f_1,f_0,M_\alpha} (ij) = W_{ij}, \\
W_{M_{\alpha},f_0,f_1}(ij) \pm W_{M_{\alpha},f_0,M_\beta} (ij) = \pm \delta_{\alpha\beta} W_{ij}, \\
W_{f_0,f_0,f_1}(ij) \pm W_{f_0,f_0,M_\alpha} (ij) = -W_{ij}.
\end{array}
\end{equation}
Equations (\ref{e1}) have the following solution:
\begin{equation} \label{e1}
\begin{array}{l}
W_{f_1,f_0,f_1}(ij) = W_{ij}, \quad W_{f_1,f_0,M_\alpha} = 0;
\\
W_{M_{\alpha},f_0,f_1} = 0, \quad W_{M_{\alpha},f_0,M_\beta} = W_{ij};
\\
W_{f_0,f_0,f_1}(ij) = -W_{ij}, \quad W_{f_0,f_0,M_\alpha} = 0.
\end{array}
\end{equation}
This gives us 20 (from 216) elements $W_{x,y,z}$.

 Similarly we can describe other ``simple" cases: tunneling a form double-occupied site to a free one, tunneling from a single-occupied site to another single-occupied one and so on. Each case gives us a set of equations. As a result using these equations one can find all the hopping rates $W_{xyz}$.

\section{Appendix 2: Numerical technics}

Our computations are based on the rate equation (\ref{Spin-rel-gen}) considering $M_{\alpha}^{(k)}(t)$ --- the on-site magnetic moments projections averaged with time-dependent density matrix. Here index $\alpha$ corresponds to a cartesian coordinate and $k$ numerates the sites.

We start with the equilibrium state (i.e. the equilibrium density matrix) of the system. At the time $t=0$ small non-equilibrium part is added to the density matrix so that each site has higher probability to be single-occupied with electron with spin up than to be single occupied with an electron with spin down. It correspond to the appearance of the small $z$-component of the magnetic moment on each site $M_{z}^{(0)}(0) = M_{z,0} \ll 1$. The consequent relaxation follows the kinetic equation that relate different magnetizations $M_{\alpha}^{(k)}$ between themselves.

In our simulation we consider the numerical samples containing $10^4$ sites with random (uncorrelated) positions inside a cube with the side $L$. We apply periodic boundary conditions. The equation (\ref{Spin-rel-gen}) is considered in a matrix form $d m_i /dt = \sum_{j}T_{ij}m_j$. Here $m_i$ correspond to the relative magnetizations $m_i(t) = M_i(t)/M_i(0)$. Indexes $i$ and $j$ stand for both the number of site and cartesian coordinate corresponding to the spin projection ($x$,$y$ or $z$).
The equation is solved with simple iterations $m_{i}(t + \delta t) = m_{i}(t) + \delta t T_{ij}m_j(t)$. The step $\delta t$ was different in different numerical experiment but was selected to be at least 10 times smaller than the spin precession time in a random magnetic field and the minimal possible spin transition time $\Upsilon_0^{-1}$. The results were averaged over at least 100 disorder realizations (i.e. different realizations of random fields and in the case of random site distribution --- different positions of sites).

To facilitate the calculation we used a cutoff for the hopping length. The hopping rates exponentially depend on the hopping length therefore very long hops do not contribute significantly to the system dynamics.  So we consider only the hops which are not longer then some critical distance. This distance is selected in such a way that each site have in average 20 neighbors which are included into the computation. We have checked that the results do not depend on the cut-off distance.


\begin{thebibliography}{99}
\bibitem{comment-magn} Note that in our approach we consider magnetization as a part of the on-site density matrix. It can be
understood as the ensemble-averaged on-site magnetization (one can always imagine the ensemble
of similar systems with same disorder realizations and make averaging over these systems).
As long as a site has some probability to be single-occupied it can have the magnetization.
\end{thebibliography}
\end{document}